\documentclass[prd,aps,preprintnumbers,showpacs,twocolumn,nofootinbib,superscriptaddress,notitlepage]{revtex4-1}
\usepackage[normalem]{ulem}
\usepackage{epsfig}
\usepackage{amsfonts}
\usepackage{amsmath}
\usepackage{slashed}
\usepackage{graphicx}
\usepackage{color}
\usepackage{mathtools}
\usepackage{subfigure}
\usepackage{stmaryrd}
\usepackage{amssymb,psfrag}
\usepackage[normalem]{ulem}
\usepackage{caption}

\newcommand{\beq}{\begin{eqnarray}}
\newcommand{\eeq}{\end{eqnarray}}

\newcommand{\nn}{\nonumber}

\DeclareRobustCommand{\fig}[1]{Fig.~\ref{fig:#1}}

\begin{document}

\title{Exploring the neutron momentum distribution in nuclei through $\gamma n \to \pi^- p$ at an electron-positron collider}

\author{Zi-Wei Yan}
\email{yanziwei@stu.zzu.edu.cn}
\affiliation{Key Laboratory for Particle Astrophysics and Cosmology, Ministry of Education (MoE), School of Physics and Astronomy, Shanghai Jiao Tong University, Shanghai 201210, China}
\affiliation{School of Physics, Zhengzhou University, Zhengzhou, Henan 450001, China}

\author{Shu-man Hu}
\email{hushm2025@lzu.edu.cn, corresponding author}
\affiliation{School of Nuclear Science and Technology, Lanzhou University, Lanzhou 730000, China}

\author{Wei Wang}
\email{wei.wang@sjtu.edu.cn, corresponding author}
\affiliation{State Key Laboratory of Dark Matter Physics, Shanghai Key Laboratory for Particle Physics and Cosmology,
Key Laboratory for Particle Astrophysics and Cosmology (MOE),
School of Physics and Astronomy, Shanghai Jiao Tong University, Shanghai 200240, China}
\affiliation{Southern Center for Nuclear-Science Theory (SCNT), Institute of Modern Physics,
Chinese Academy of Sciences, Huizhou 516000, Guangdong Province, China}

\author{Ji Xu}
\email{xuji@lzu.edu.cn, corresponding author}
\affiliation{School of Nuclear Science and Technology, Lanzhou University, Lanzhou 730000, China}
\affiliation{Key Laboratory for Particle Astrophysics and Cosmology, Ministry of Education (MoE), School of Physics and Astronomy, Shanghai Jiao Tong University, Shanghai 201210, China}

\author{Fu-Sheng Yu}
\email{yufsh@lzu.edu.cn, corresponding author}
\affiliation{School of Nuclear Science and Technology, Lanzhou University, Lanzhou 730000, China}

\author{Ya-Teng Zhang}
\email{zhangyateng@zzu.edu.cn, corresponding author}
\affiliation{School of Physics, Zhengzhou University, Zhengzhou, Henan 450001, China}

\begin{abstract}
The neutron momentum distribution is essential both for reliably extracting fundamental free neutron observables from nuclear measurements and for probing the tensor force via the high-momentum neutron fraction, which is crucial to the theoretical understanding of short-range correlations (SRCs). In this work, we investigate this distribution by studying the $\gamma n \to \pi^- p$ process at an electron-positron collider, proposing to utilize the beryllium beam pipe at the Beijing Spectrometer III (BESIII). The cross sections for this process on both deuteron and beryllium targets are calculated within the impulse approximation framework. We also evaluate the effective luminosity of the photon flux from radiative Bhabha scattering, taking into account the distribution of target materials within the BESIII experimental setup. Our results show that tens of thousands of events can be generated at BESIII, offering the potential for precise measurements of the neutron momentum distribution. These findings suggest that electron-positron colliders could play a valuable role in elucidating nuclear structure and advancing our understanding of nonperturbative QCD, offering promising new avenues for both particle and nuclear physics.
\end{abstract}

\maketitle

\section{Introduction}
\label{sec:introduction}

The strong interaction binds protons and neutrons---collectively known as nucleons---into stable atomic nuclei, emerging from the intricate dynamics of quarks and gluons governed by quantum chromodynamics (QCD). Understanding the motion of nucleons in the nuclear medium is a central objective of nuclear physics. It is also essential for the quantitative interpretation of reactions involving nuclear targets and has broad implications for the description of nuclear matter \cite{Frankfurt:1988nt,Hen:2016kwk,Arrington:2022sov}.

On the one hand, the extraction of elementary cross sections on the neutron from nuclear data requires a precise knowledge of the bound-nucleon momentum distribution in order to disentangle nucleon-motion effects from the observables of interest \cite{CiofidegliAtti:1995qe,Fujii:2001rc,CiofidegliAtti:2007ork,Das:2023vvn}. On the other hand, ab initio variational Monte Carlo (VMC) calculations for light nuclei reveal that the single-nucleon momentum distributions exhibit a universal high-momentum tail, which is predominantly generated by the tensor component of the nucleon-nucleon interaction, establishing these distributions as a direct probe of short-range correlations \cite{Wiringa:2013ala,Piarulli:2022ulk,Datar:2013pbd,Carlson:2014vla,Piarulli:2017dwd}. Among them, the neutron momentum distribution plays a particularly critical role. The absence of free neutron targets makes deuteron-based measurements the primary source of neutron-specific observables, and the accuracy of the extracted neutron cross sections hinges directly on the neutron momentum distribution employed in the nuclear corrections. Moreover, in neutron-rich nuclei the fraction of high-momentum neutrons is substantially smaller than that of high-momentum protons, a consequence of the tensor force acting predominantly in isoscalar $np$ pairs \cite{Hen:2014nza,CLAS:2020mom,CLAS:2019vsb,Subedi:2008zz}. Reliable knowledge of the neutron momentum distribution is therefore essential for extracting neutron-target observables from nuclear data, in particular the electromagnetic transition amplitudes of nucleon resonances excited from the neutron, and for elucidating the isospin structure of nucleon-nucleon correlations \cite{Krusche:2003ik,Kamano:2016bgm}.

The momentum distribution of nucleons in a nucleus can be divided into two distinct regimes. Below the Fermi momentum $k_F$, the motion is dominated by Fermi motion---a quantum-mechanical consequence of the nuclear mean field---while above $k_F$, it is governed by SRCs. These SRCs arise from the strong short-distance interaction between nucleons, producing pairs with large relative momentum and small center-of-mass momentum. They are regarded as a key bridge connecting quark-gluon and nucleonic degrees of freedom \cite{Frankfurt:1993sp,Arrington:2011xs,Miller:2020eyc,Zhang:2025nst}, and recent studies have linked them closely to the well-known EMC effect \cite{EuropeanMuon:1983wih,Weinstein:2010rt,Hen:2012fm,nCTEQ:2023cpo,Hu:2026mmr,Wang:2025mur,Courtoy:2025ppd,Paakkinen:2025pcw,Yang:2024bzq,Huang:2025kmd,Wang:2024cpx,Mirjalili:2025yxe,Cai:2025txx,Yang:2023zmr,Huang:2021cac}. However, existing experimental data remain insufficient, and further measurements are needed to fully characterize SRCs.

In recent years, it has been proposed to utilize the beam pipe at the Beijing Spectrometer III (BESIII) as a nuclear target for studying nucleon structure \cite{Yuan:2021yks,Dai:2022wpg}. Since this proposal, studies utilizing the beam pipe at BESIII have attracted considerable attention \cite{Xu:2024pih,BESIII:2023clq,BESIII:2024geh,BESIII:2026iti,BESIII:2025yup}. The beam pipe is primarily composed of beryllium ($^9$Be), and several measurements have already been performed using this target. However, no study to date has focused on studying the nuclear physics through reactions with the beam pipe. In Ref.\,\cite{Wang:2024ikx}, it was proposed to use sub- and near-threshold $\phi$-meson photoproduction at BESIII to investigate nucleon structure, but the estimated yield is limited to only about ten events, making a precise measurement unfeasible. Therefore, a process capable of producing substantially more events is imperative.

Pion photoproduction measurements facilitate the understanding of the strong force in the low-energy regime. Most of the experimental efforts over the last few decades have focused on pion production from proton targets, namely $\gamma p \to \pi^0 p$ and $\gamma p \to \pi^+ n$ \cite{Ireland:2019uwn,A2:2019yud}. In contrast, pion photoproduction off neutrons has received relatively little attention. This is understandable, as there is no free neutron target available. Currently, the primary approach is to extract the $\gamma n \to \pi^- p$ process from data on pion photoproduction on the deuteron \cite{Strandberg:2018djk,Briscoe:2021siu}. In this work, we propose that the $\gamma n \to \pi^- p$ process can serve as an ideal probe for studying the neutron momentum distribution in nuclei. We aim to leverage the produced continuous photon beam with feasible energies at BESIII, along with its high luminosity and high-performance detectors, to their full potential. The $\gamma n \to \pi^- p$ process possesses a typical cross section ($100\,\rm{\mu b}$) that is approximately three orders of magnitude larger than that of near-threshold $\phi$-meson photoproduction \cite{Briscoe:2020qat}, making it an ideal candidate for probing the neutron momentum distribution with high statistical precision. Additionally, we explore the effective luminosity of the photon flux produced by radiative Bhabha scattering, considering the distribution of target materials within the experimental setup at BESIII \cite{BESIII:2009fln,BESIII:2020nme}. We find that the calculated number of events can reach tens of thousands, demonstrating that BESIII can perform very precise measurements of this process. Future experimental efforts to validate this proposal have the potential to advance our understanding of nuclear structure significantly.

This paper is organized as follows. In Sec.\ref{sec:theoretical}, we present the theoretical framework for calculating the $\gamma n \to \pi^- p$ cross section on bound neutrons, including the impulse approximation, the MAID model for the free cross section, and the neutron momentum distributions for both deuteron and beryllium. In Sec.\,\ref{sec:besiii}, we present the opportunities at BESIII, including discussion of the photon source from radiative Bhabha scattering and the estimation of event numbers. Sec.\,\ref{sec:summary} is reserved for summary and outlook.

\section{Theoretical analysis}
\label{sec:theoretical}
\begin{figure}[htbp]
\centering
\includegraphics[width=1.00\columnwidth]{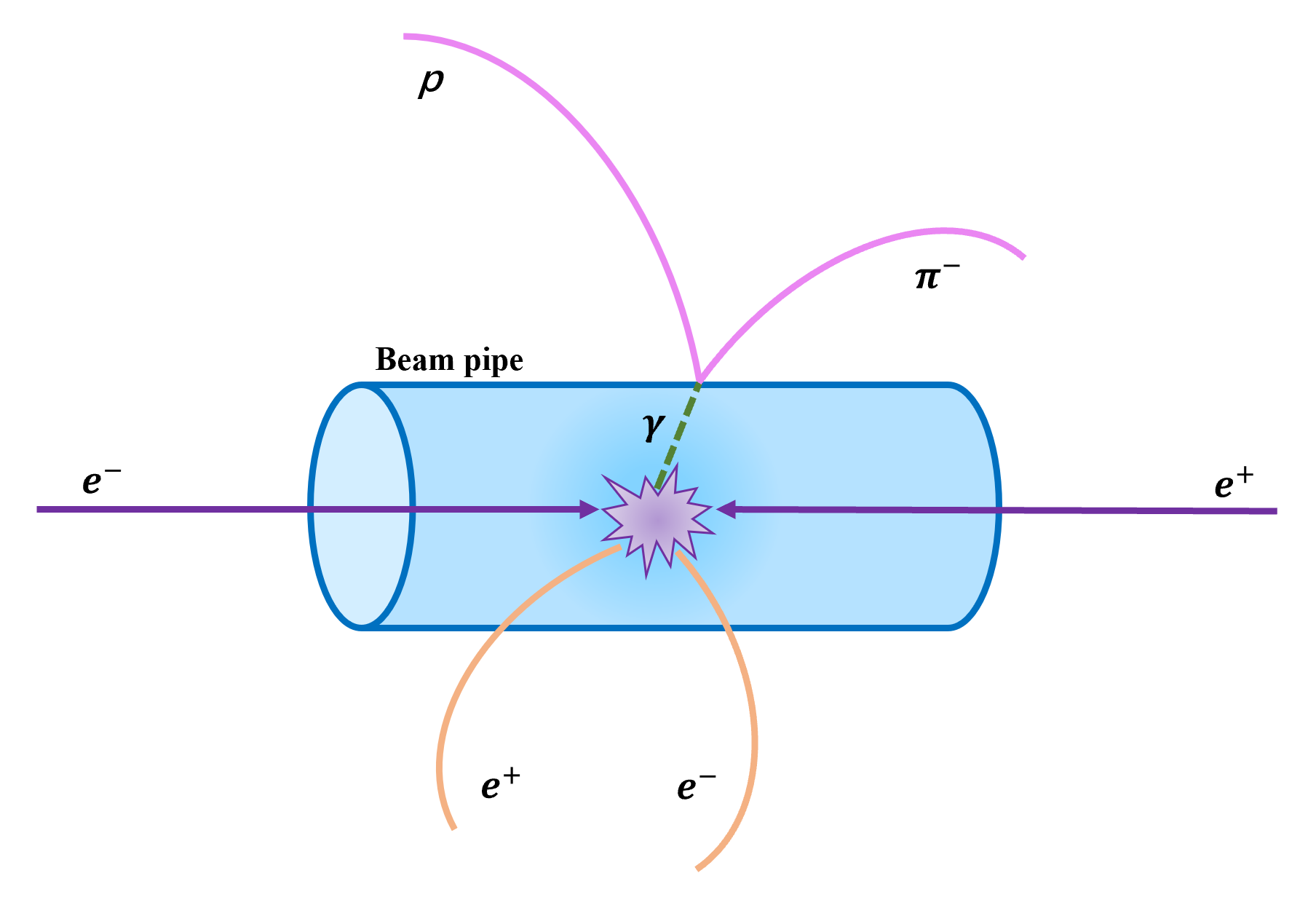}
\caption{Schematic diagram of the $\gamma n \to \pi^- p$ process, where the neutron is bound in the nucleus. The photon interacts with the bound neutron, producing a $\pi^-$ meson and a proton in the final state.}
\label{fig:schematic}
\end{figure}

The process of interest is illustrated in \fig{schematic}. A photon with four-momentum $k_\gamma = (E_\gamma, \vec{\bf{k}}_\gamma)$ interacts with a neutron bound in a nucleus, where the neutron carries four-momentum $p_n = (E_n, \vec{\bf{p}}_n)$. The final state consists of a $\pi^-$ with momentum $k_\pi = (E_\pi, \vec{\bf{k}}_\pi)$ and a proton with momentum $p_p = (E_p, \vec{\bf{p}}_p)$. Here, the initial state of the neutron is the subject of our interest.

If the initial neutron were free, its experimentally measurable momentum would lie on the mass shell, $E_n^2 - |\vec{\bf{p}}_n|^2 = m_n^2$. However, when the neutron is bound inside a nucleus, it is in motion due to both Fermi motion and SRCs. We can define an experimentally measurable quantity: the missing energy $E_{\textrm{miss}} = m_n - E_n$, which quantifies the difference between the neutron's rest mass and its energy, thereby characterizing its off-shell nature.

When the neutron is bound within a nucleus, the $\gamma n \to \pi^- p$ process provides a sensitive probe for studying neutron momentum distributions. In this scenario, the finite momentum distribution and off-shellness of the bound neutron imply that the experimentally measured cross section is modified compared to the free case. Here we adopt the impulse approximation (IA) framework \cite{Hatta:2019ocp,Xu:2019wso}. In this approximation, the photon interacts with only one nucleon inside the nucleus---specifically the neutron---while the remaining nuclear system acts as a spectator and does not participate in the reaction. Under this assumption, the cross section of the process is given by a convolution of the free cross section with the neutron momentum distribution inside the nucleus,
\beq\label{eq:sigmabound}
   \sigma_{\textrm{bound}}(E_\gamma) &=& \int \frac{d^3 \vec{\bf{p}}_n}{(2\pi)^3} \, \sigma_{\textrm{free}}(\sqrt{s}) \, \rho_n(|\vec{\bf{p}}_n|) \,,
\eeq
which can be further simplified as:
\beq\label{eq:sigmabound}
   \sigma_{\textrm{bound}}(E_\gamma) &=& \frac{1}{(2\pi)^2} \int_0^\infty d|\vec{\bf{p}}_n| |\vec{\bf{p}}_n|^2 \nn\\
    &&\times \int_0^\pi d\theta \, \sin\theta \, \sigma_{\textrm{free}}(\sqrt{s}) \, \rho_n(|\vec{\bf{p}}_n|) \,,
\eeq
where $\sigma_{\textrm{free}}(\sqrt{s})$ is the free $\gamma n \to \pi^- p$ cross section, and $\theta$ is the angle between the momenta of the initial photon and the bound neutron. The center-of-mass energy squared $s$ of the $\gamma n$ system is given by
\beq\label{eq:sqrts}
  s &=& m_n^2 + E_{\textrm{miss}}^2 - 2 E_{\textrm{miss}} (m_n + E_\gamma) + 2m_n E_\gamma \nn\\
  && - |\vec{\bf{p}}_n|^2 - 2 E_\gamma |\vec{\bf{p}}_n| \cos\theta \,.
\eeq
The missing energy $E_{\textrm{miss}}$ depends on the neutron momentum. For nearly all non-SRC neutrons, which can be treated as approximately on-shell, one has $E_{\textrm{miss}} \simeq 0$. For neutrons in SRC pairs, which carry high momenta and are significantly off-shell, a linear relationship between $|\vec{\bf{p}}_n|$ and $E_{\textrm{miss}}$ is expected based on electron-nucleon scattering data \cite{CLAS:2020mom}. We adopt the following functional form:
\beq\label{eq:Emiss}
  E_{\textrm{miss}}(|\vec{\bf{p}}_n|)=\left\{\begin{array}{cc}
0 \,, & \quad  |\vec{\bf{p}}_n|<k_F \\ \\
0.446 \, |\vec{\bf{p}}_n| - 0.098 \, \textrm{GeV} \,, & \quad  |\vec{\bf{p}}_n|>k_F
\end{array}\,.\right.
\eeq
In realistic nuclear structure theory, the typical missing energy for a mean-field neutron is on the order of $E_{\textrm{miss}}\sim 10$\,MeV. In the present analysis, such effects are neglected. This simplification is not expected to significantly affect the quantities of primary interest here. A more rigorous treatment would employ the full energy-momentum spectrum from ab initio nuclear calculations. However, we are more interested in the high-momentum regime, where Ref.\,\cite{CLAS:2020mom} indicates an approximately linear relationship between $|\vec{\bf{p}}_n|$ and $E_{\textrm{miss}}$. This simple form gives a transparent physical picture, whether this picture can be validated by future measurements at BESIII is one of the key questions we wish to address.

\subsection{The free $\gamma n \to \pi^- p$ cross section}
\label{sec:freesigma}

Let us begin by discussing the free cross section $\sigma_{\textrm{free}}(E_\gamma)$ in Eq.\,(\ref{eq:sigmabound}). Measurements of pion photoproduction on both proton and neutron targets have a very long history \cite{White:1952zz}. However, since there is no free neutron target available, the $\gamma n \to \pi^- p$ process can only be accessed indirectly through measurements on the deuteron, i.e., through the reaction $\gamma d \to \pi^- p p$. For the energy range from threshold up to the resonance region, theoretical predictions can be provided by unitary isobar models such as MAID \cite{Drechsel:2007if,Hilt:2013fda}.

We adopt the MAID (Mainz Unitary Isobar Model) program to provide the free $\gamma n \to \pi^- p$ cross section. MAID is a unitary isobar model within a dispersion-relation framework, designed for the systematic description of single-pion photo- and electroproduction on the nucleon. It thereby offers a unified and readily accessible interface for the theoretical interpretation of experimental data and the generation of predictions. The MAID model is well suited for this task for several reasons. First, it is built upon rigorous constraints from unitarity and analyticity, with multipoles that respect low-energy theorems in the threshold region and can be naturally extrapolated into the resonance region, ensuring a reliable description from near-threshold to intermediate energies. Second, although direct experimental data for the neutron-target channel $\gamma n \to \pi^- p$ are extremely scarce, the MAID model provides a consistent description of this process that respects unitarity and is compatible with existing constraints. Third, the reaction amplitude in the threshold region is dominated by the $E_{0+}$ multipole, where MAID incorporates essential constraints from chiral perturbation theory, while at higher energies up to $E_\gamma \sim 500\,\textrm{MeV}$ it includes the $\Delta(1232)$ resonance and higher partial waves, preserving a physically coherent description across the full energy range.

\begin{figure}[htbp]
\centering
\includegraphics[width=1.00\columnwidth]{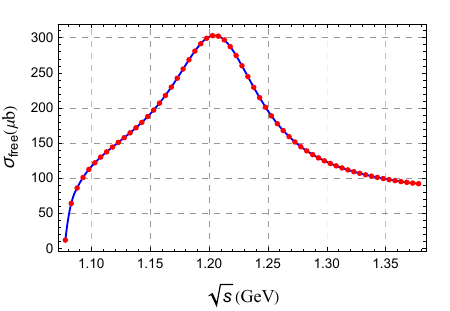}
\caption{The total cross section of $\gamma n \to \pi^- p$ as a function of center-of-mass energy $\sqrt{s}$, provided by MAID. Here the neutron in the initial state is a ``free'' neutron. The red dotted points represent the MAID data, while the blue solid curve is an interpolation.}
\label{fig:freesigma}
\end{figure}

Fig.\,\ref{fig:freesigma} shows the cross section for the $\gamma n \to \pi^- p$ reaction from threshold up to $1.35\,\textrm{GeV}$, as provided by the MAID model. A prominent peak around $\sqrt{s} \sim 1.20\,\textrm{GeV}$ corresponds to the excitation of the $\Delta(1232)$ resonance.

\subsection{Neutron momentum distributions in nuclei}
\label{sec:momdist}

We then move on to the neutron momentum distribution. In the atomic nucleus, nucleons are not stationary but in constant motion, and thus possess momenta. One of the most common types of motion is Fermi motion. For beryllium $^{9}$Be, the Fermi momentum is taken to be $k_F = 206\,\textrm{MeV}$ \cite{Moniz:1971mt,Vanhalst:2012ur}. This value does not distinguish between neutrons and protons, as the Fermi motion is taken to be the same for both. Below the Fermi momentum, the nucleon momentum distribution is dominated by the nuclear mean field and can be described by mean-field wave-function calculations. Another kind of motion comes from the short-range nucleon-nucleon structures, in which the participating nucleons form predominantly neutron-proton pairs with relative momenta several times the Fermi momentum \cite{Hen:2014nzasupp}.

Both of these motions influence the nucleon momentum distributions in nuclei, with their relative contributions varying across nuclear species. In the following, we shall adopt two distinct formalisms to describe the neutron momentum distributions in beryllium and deuterium.

For the neutron momentum distribution in beryllium, we utilize the np-SRC dominance model \cite{Hen:2014nza,Sargsian:2012sm}. It should be emphasized that the full momentum distribution cannot be directly observed in experiments. This model describes the nucleon momentum distribution as two components: an independent-nucleon distribution up to a transition momentum, typically taken as the Fermi momentum $k_F$, and a high-momentum tail above it dominated by SRC pairs. The resulting neutron momentum distribution is given by:
\beq\label{eq:rhomodel}
  \rho_n(|\vec{\bf{p}}_n|)=\left\{\begin{array}{cc}
\eta \cdot \rho_n^{\textrm{MF}}(|\vec{\bf{p}}_n|) \,, & \quad  |\vec{\bf{p}}_n|<k_F \\[0.5em]
\displaystyle\frac{A}{2 N} \cdot a_2^n(A/d) \cdot \rho_d(|\vec{\bf{p}}_n|) \,, & \quad  |\vec{\bf{p}}_n|>k_F
\end{array}\,,\right.
\eeq
which is isotropic (do not depend on the direction of $\vec{\bf{p}}_n$) and obeys the normalization
\beq\label{eq:normalization}
  \int \frac{d^3 \vec{\bf{p}}_n}{(2\pi)^3} \, \rho_n(|\vec{\bf{p}}_n|) = 1 \,.
\eeq
In Eq.\,(\ref{eq:rhomodel}), the $\rho_n(|\vec{\bf{p}}_n|)$ is the neutron momentum distribution in nucleus A, $N$ is the number of neutrons, and $A$ is the mass number. The ingredients and approximations used in this model are as follows.

\begin{itemize}
  \item $\rho_n^{\textrm{MF}}(|\vec{\bf{p}}_n|)$ is the mean-field neutron momentum distribution, for which we adopt the Boltzmann parametrization from the Woods-Saxon model \cite{Vanhalst:2012ur},
      \begin{eqnarray}
        \rho_p^{\mathrm{MF}}(|\vec{\bf{p}}_n|)=\left(\dfrac{2\pi}{m_Np_T}\right)^{3/2} e^{-\frac{|\vec{\bf{p}}_n|^2}{2m_N p_T}} \,.\nn
      \end{eqnarray}
      For $^9$Be, the Boltzmann parameter is determined to be $p_T = 11.8\,\textrm{MeV}$.

  \item $\rho_d(|\vec{\bf{p}}_n|)$ is the neutron momentum distribution in the deuteron. The high-momentum tail ($|\vec{\bf{p}}_n| > k_F$) is taken from the VMC calculation with the AV18+UX potential \cite{Wiringa:2013ala}. The second line in Eq.\,(\ref{eq:rhomodel}) indicates that in the range of $|\vec{\bf{p}}_n| > k_F$, the momentum distribution is defined by $np$ correlations only---this is the essence of the np-SRC dominance model.

  \item $a_2^n(A/d)$ is the SRC scaling factor, defined as the experimentally determined per-neutron probability of finding a high-momentum neutron in nucleus A relative to the deuteron. We take $a_2^n(^9\textrm{Be}/d) = 3.52$ from the updated CLAS data \cite{CLAS:2019vsb}.

  \item $k_F = 206\,\textrm{MeV}$ is the Fermi momentum, taken as the transition momentum in this model.

  \item $\eta$ is a normalization factor chosen such that $\rho_n(|\vec{\bf{p}}_n|)$ satisfies Eq.\,(\ref{eq:normalization}), which gives $\eta = 1.063$.
\end{itemize}

In addition to beryllium, we are also interested in the $\gamma n \to \pi^- p$ process on a neutron bound in the deuteron, because there exist recent measurements of the reaction $\gamma d \to \pi^- p p$ in the threshold region \cite{Strandberg:2018djk}. These data from the PIONS@MAX-lab Collaboration provide a valuable benchmark for our theoretical predictions.

For the neutron momentum distribution in the deuteron, we adopt the results of ab initio VMC calculations with the AV18+UX potential \cite{Wiringa:2013ala}. The wave functions have been generated for a Hamiltonian containing the Argonne V$_{18}$ two-nucleon and Urbana X three-nucleon potentials. These calculations are among the most reliable for light nuclei, providing momentum distributions that are fully consistent both with the known properties of light nuclei and with recent results from nuclear lattice effective field theory (NLEFT) \cite{Lee:2008fa,Shen:2024qzi,Wu:2026tis}.

\subsection{The bound $\gamma n \to \pi^- p$ cross section}
\label{sec:boundsigma}

\subsubsection{$\gamma n \to \pi^- p$ with neutron from deuteron}
\label{sec:deuteronresult}

We first consider the case where the photon interacts with a neutron bound in the deuteron. The overall reaction is $\gamma d \to \pi^- p p$. Consequently, the cross section should incorporate a symmetry factor of $I = 1/2$ arising from the two identical protons in the final state. This factor is essential for a proper comparison with the experimental data.

We note that the threshold energy for $\gamma n \to \pi^- p$ on a free neutron is $E_\gamma^{\textrm{th}} = 0.148\,\textrm{GeV}$. However, since the neutron in the deuteron is not at rest, the reaction can in principle proceed at photon energies below this threshold, although the cross section is typically very small in this sub-threshold region.

\begin{figure}[htbp]
\centering
\includegraphics[width=1.00\columnwidth]{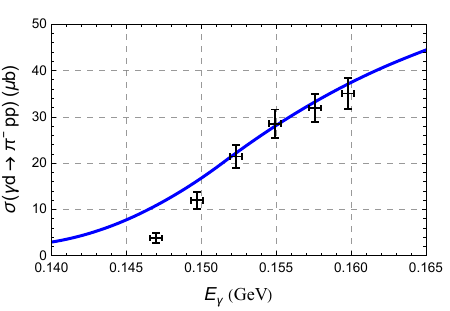}
\caption{The cross section of the reaction $\gamma d \to \pi^- p p$ in the near-threshold region, derived from Eq.\,(\ref{eq:sigmabound}) with the AV18+UX deuteron momentum distribution (blue solid line). The black points with error bars denote the PIONS@MAX-lab data.}
\label{fig:CSDeuteronThreshold}
\end{figure}

\begin{figure}[htbp]
\centering
\includegraphics[width=1.00\columnwidth]{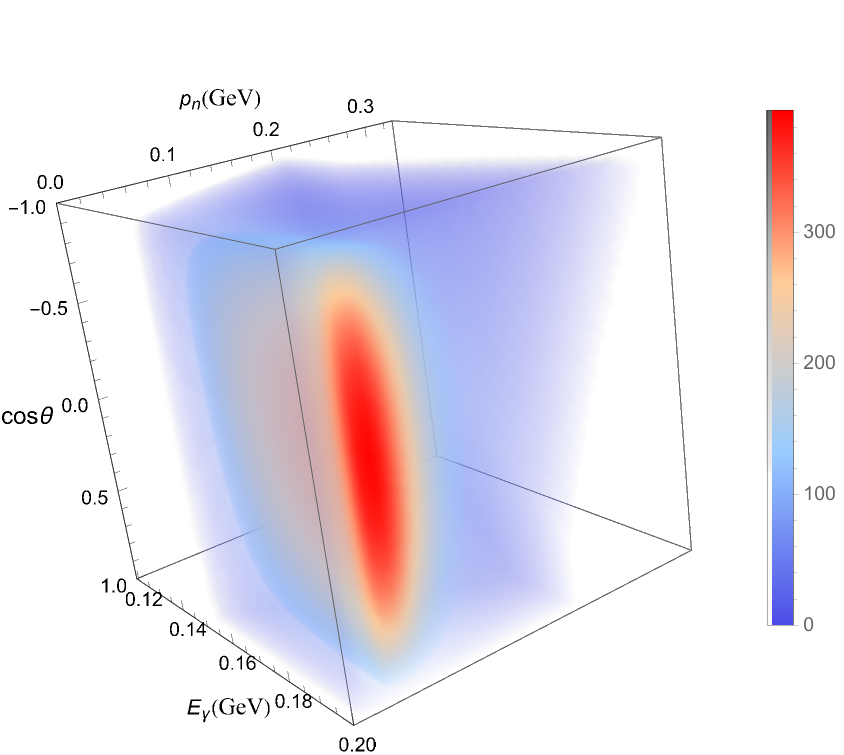}
\caption{The integrand in Eq.\,(\ref{eq:sigmabound}) displayed in the three-dimensional phase space spanned by $\cos\theta$, $|\vec{\bf{p}}_n|$, and $E_\gamma$, for the case of $\gamma n \to \pi^- p$ with the neutron from the deuteron.}
\label{fig:DeuteronThreshold3D}
\end{figure}

Fig.\,\ref{fig:CSDeuteronThreshold} shows the bound cross section $\sigma_{\textrm{bound}}(E_\gamma)$ for the reaction $\gamma d \to \pi^- p p$ in the near-threshold region (0.135\,--\,0.170\,GeV), calculated using Eq.\,(\ref{eq:sigmabound}) with the AV18+UX deuteron momentum distribution and the symmetry factor $I = 1/2$. For a more detailed analysis of the contributions to $\sigma_{\textrm{bound}}(E_\gamma)$, the integrand in Eq.\,(\ref{eq:sigmabound}) is mapped onto the three-dimensional phase space spanned by $\cos\theta$, $|\vec{\bf{p}}_n|$, and $E_\gamma$, as illustrated in \fig{DeuteronThreshold3D}.

It can be observed that in the near-threshold region, only neutrons with momenta $|\vec{\bf{p}}_n| < 0.3\,\textrm{GeV}$ contribute to the total cross section, while the high-momentum component does not contribute. This imposes a strong constraint on the experimental detection of high-momentum neutrons near the threshold. In the near-threshold region, the result obtained using the IA framework together with the AV18+UX deuteron momentum distribution aligns with the PIONS@MAX-lab data. The agreement could be further improved by incorporating final state interactions (FSI) as a correction to the impulse approximation \cite{Briscoe:2020qat}. However, the current precision is already sufficient for our purpose of estimating the event yield for the $\gamma n \to \pi^- p$ process at BESIII, and a more refined theoretical treatment will be pursued in future work.

\subsubsection{$\gamma n \to \pi^- p$ with neutron from beryllium}
\label{sec:berylliumresult}

\begin{figure}[htbp]
\centering
\includegraphics[width=1.00\columnwidth]{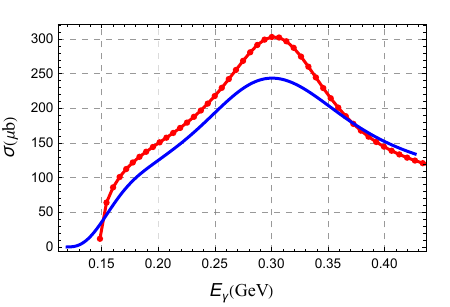}
\caption{The cross section of the reaction $\gamma n \to \pi^- p$ where the initial neutron is bound in $^9$Be. The red dotted line represents the free cross section $\sigma_{\textrm{free}}(E_\gamma)$ given by the MAID model. The blue solid line is the bound cross section $\sigma_{\textrm{bound}}(E_\gamma)$ derived from Eq.\,(\ref{eq:sigmabound}), in which the neutron momentum distribution is given by Eq.\,(\ref{eq:rhomodel}).}
\label{fig:CSBeryllium}
\end{figure}

Using the neutron momentum distribution $\rho_n(|\vec{\bf{p}}_n|)$ given in Eq.\,(\ref{eq:rhomodel}), we calculate the cross section for the reaction $\gamma n \to \pi^- p$ on a neutron bound in beryllium. The results are shown in Fig.\,\ref{fig:CSBeryllium}. It can be seen that in the range of $E_\gamma$ from threshold up to about $0.5\,\textrm{GeV}$, the nuclear effects arising from the neutron's motion inside the nucleus significantly modify the cross section. In particular, near $E_\gamma \sim 0.3\,\textrm{GeV}$, the cross section differs by approximately $50\,\mu\textrm{b}$ between the free neutron and the bound neutron cases.

\begin{figure}[htbp]
\centering
\includegraphics[width=1.00\columnwidth]{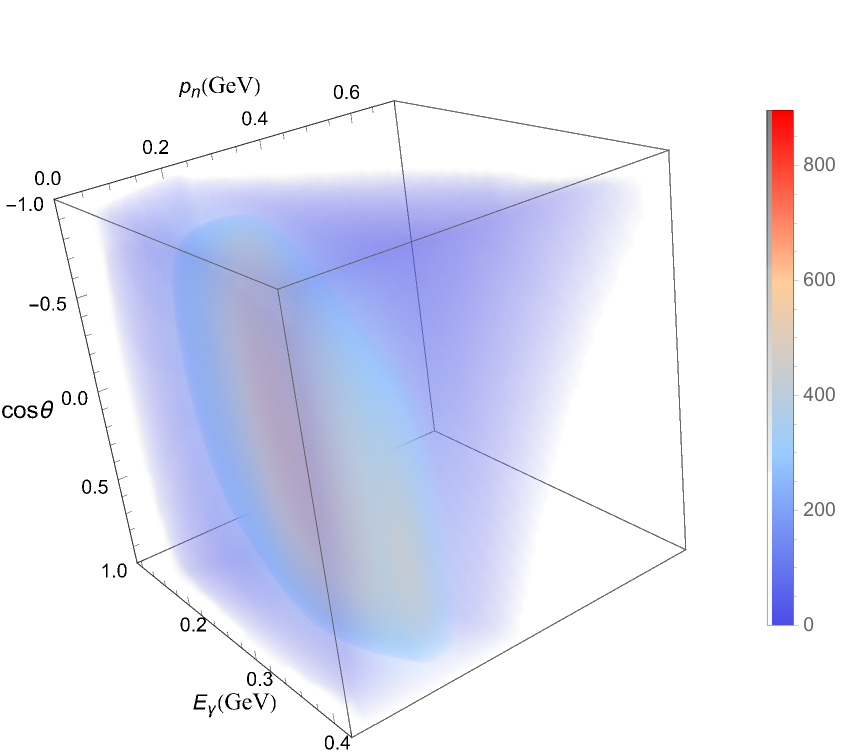}
\caption{The integrand in Eq.\,(\ref{eq:sigmabound}) displayed in the three-dimensional phase space spanned by $\cos\theta$, $|\vec{\bf{p}}_n|$, and $E_\gamma$, for the case of $\gamma n \to \pi^- p$ with the neutron from $^9$Be. Both low-momentum (Fermi motion) and high-momentum (SRC) neutrons contribute to the bound cross section.}
\label{fig:Beryllium3D}
\end{figure}

Similarly, Fig.\,\ref{fig:Beryllium3D} displays the integrand in Eq.\,(\ref{eq:sigmabound}) mapped onto the three-dimensional phase space for the beryllium case. One can observe that in the photon energy range from threshold to $0.4\,\textrm{GeV}$, both low-momentum ($|\vec{\bf{p}}_n| < k_F$) and high-momentum ($|\vec{\bf{p}}_n| > k_F$) neutrons contribute to the total cross section. This is in contrast to the deuteron case, where the high-momentum component is negligible. The pronounced contribution from high-momentum neutrons in $^9$Be is a direct consequence of the np-SRC dominance model, which predicts an enhanced high-momentum tail due to the scaling factor $a_2^n(^9\textrm{Be}/d) = 3.52$.

The values of the bound cross section $\sigma_{\textrm{bound}}(E_\gamma)$ for different photon energy bins are collected in Table~\ref{tab:CSvalues}. These values will be used in the event estimation in the next section.

\begin{table}[!htbp]
\centering
\renewcommand{\arraystretch}{1.5}
\caption{The bound cross section $\sigma_{\textrm{bound}}(E_\gamma)$ for $\gamma n \to \pi^- p$ with the neutron from $^9$Be, calculated in the np-SRC dominance model.}\label{tab:CSvalues}
\begin{tabular}{c||c}
\hline\hline
 ~~~~~~~~~~$E_\gamma$(GeV)~~~~~~~~~~ & ~~~~~~~~~~$\sigma_{\textrm{bound}}(\mu\textrm{b})$~~~~~~~~~~  \\
\hline
~~~[0.10, 0.15]~~~ & ~~~$0.56$~~~  \\
~~~[0.15, 0.20]~~~ & ~~~$91.08$~~~  \\
~~~[0.20, 0.25]~~~ & ~~~$156.01$~~~  \\
~~~[0.25, 0.30]~~~ & ~~~$228.61$~~~  \\
~~~[0.30, 0.35]~~~ & ~~~$231.67$~~~  \\
~~~[0.35, 0.40]~~~ & ~~~$175.35$~~~  \\
~~~[0.40, 0.45]~~~ & ~~~$135.16$~~~  \\
\hline\hline
\end{tabular}
\end{table}

The difference between $\sigma_{\textrm{bound}}(E_\gamma)$ and $\sigma_{\textrm{free}}(E_\gamma)$ is clearly visible. By comparing the measured cross section shape with the theoretical prediction, one can quantitatively extract the nuclear modification effects and probe the neutron momentum distribution. This demonstrates that the $\gamma n \to \pi^- p$ process on beryllium beam pipe at BESIII offers a unique window into the investigation of nuclear structure, which will be discussed in the next section.

\section{The opportunities in BESIII}
\label{sec:besiii}
The BESIII detector records symmetric $e^+ e^-$ collisions provided by the BEPCII storage ring. The physics program of BESIII covers a wide range of topics in hadron spectroscopy, charmonium physics, charm meson decays, QCD tests, and $\tau$ physics. In addition to these well-established programs, the beam pipe at BESIII offers a unique opportunity for nuclear physics measurements. The materials in the beam pipe can be considered as targets of beryllium ($^9$Be), as detailed in Refs. \cite{BESIII:2009fln}. Furthermore, experimental research on beryllium is of great interest in its own right: $^9$Be is a light nucleus with a pronounced cluster structure, and understanding its neutron momentum distribution remains particularly valuable for constraining nuclear structure models.

The physical picture of our proposal is as follows. First, the radiative Bhabha process $e^+ e^- \to \gamma e^+ e^-$ produces a large number of photons with energies in the range suitable for pion photoproduction ($0.1$\,--\,$0.5\,\textrm{GeV}$). These photons can interact with materials in the beam pipe alongside the $e^+ e^-$ beam, offering a unique opportunity to investigate the neutron momentum distributions. Specifically, these photons, when produced, can interact with the beam pipe and have a probability of producing pions and protons through $\gamma n \to \pi^- p$. The final detectable state of the signal process is $e^+ e^- \pi^- p$; therefore the candidate events have four charged tracks with zero net charge. By measuring the final state $\pi^-$ and proton, and determining the interaction location, we can expect to obtain the shape of the cross section as a function of photon energy. These charged tracks, originating directly from the beam pipe rather than the interaction point, can be selected and identified with high efficiency \cite{BESIII:2023clq,BESIII:2024geh}.

\subsection{Photon source from radiative Bhabha scattering}
\label{sec:bhabha}
\begin{figure}[htbp]
\centering
\includegraphics[width=1.00\columnwidth]{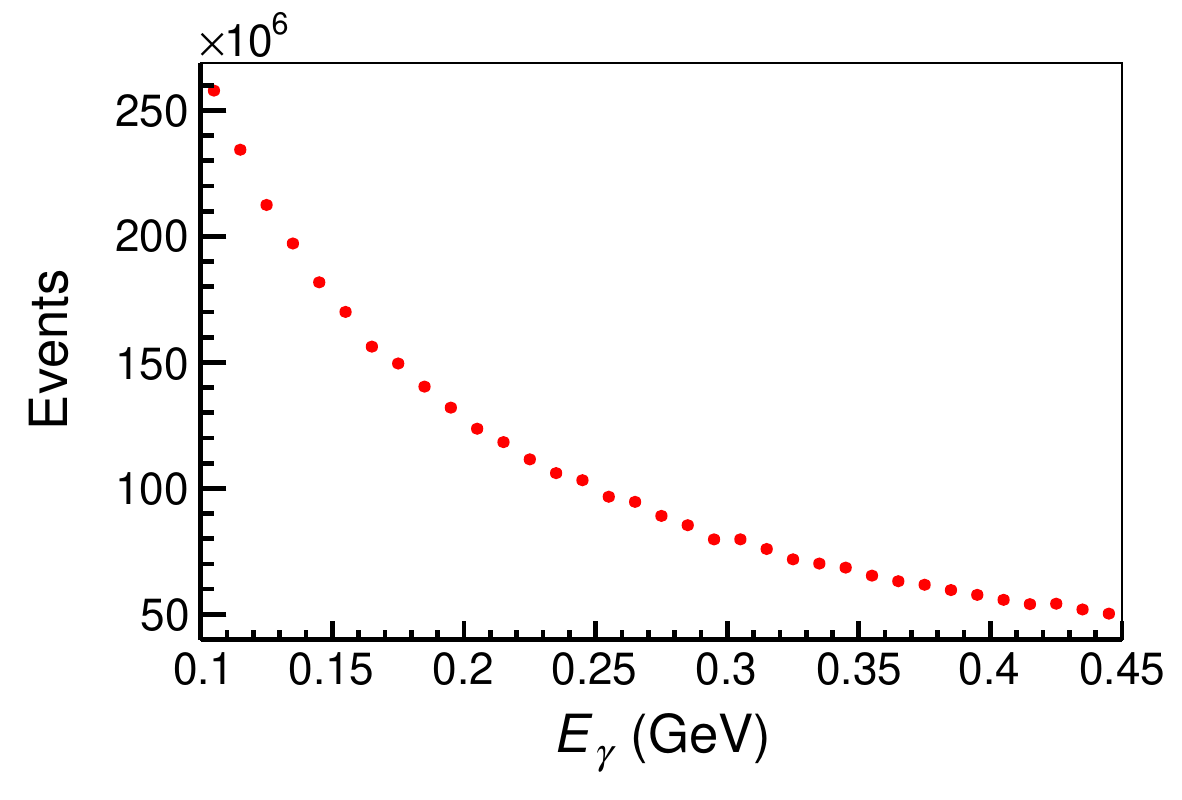}
\caption{The photon energy distribution of the radiative Bhabha process $e^+ e^- \to \gamma e^+ e^-$ at $\sqrt{s} = 3.773\,\textrm{GeV}$, simulated using the Babayaga generator. This plot is obtained from the Monte Carlo truth information. All results assume an integrated luminosity of $20\,\textrm{fb}^{-1}$.}
\label{fig:energy-Bhabha}
\end{figure}

The radiative Bhabha $e^+ e^- \to \gamma e^+ e^-$ is a well-understood QED process. Although it is typically regarded as a background in many BESIII studies \cite{Wang:2024ikx}, its large cross section makes it an ideal photon source for our purpose. Here, we aim to select the radiative Bhabha events with a single photon of energy from approximately $0.10\,\textrm{GeV}$ to $0.45\,\textrm{GeV}$, accompanied by the electron and positron in the final state recorded in the detector. To validate this method, we simulate the radiative Bhabha process using the Babayaga generator at $\sqrt{s} = 3.773\,\textrm{GeV}$ \cite{CarloniCalame:2003yt}. The photon energy distribution is illustrated in Fig.\,\ref{fig:energy-Bhabha}.

It can be observed from this figure that the radiative Bhabha process yields a very large number of photons with suitable energies, serving as an ideal photon source for pion photoproduction. We denote $\mathcal{I}_{\textrm{Bha.}}(E_\gamma)$ as the energy distribution of photon numbers produced via radiative Bhabha. The specific numerical values are listed in Table~\ref{tab:rhodistribution}, where we present the number of photons in each energy bin. One can see that the number of photons exceeds $10^8$ per $50\,\textrm{MeV}$ bin across the entire energy range of interest.

\begin{table}[!htbp]
\centering
\renewcommand{\arraystretch}{1.5}
\caption{The energy distribution of photon numbers $\mathcal{I}_{\textrm{Bha.}}(E_\gamma)$ from the radiative Bhabha process at $\sqrt{s} = 3.773\,\textrm{GeV}$, assuming an integrated luminosity of $20\,\textrm{fb}^{-1}$.}\label{tab:rhodistribution}
\begin{tabular}{c||c}
\hline\hline
 ~~~~~~~~~~$E_\gamma$(GeV)~~~~~~~~~~ & ~~~~~~~~~~$\mathcal{I}_{\textrm{Bha.}}$~~~~~~~~~~  \\
\hline
~~~[0.10, 0.15]~~~ & ~~~$1.08 \times 10^{9}$~~~  \\
~~~[0.15, 0.20]~~~ & ~~~$7.48 \times 10^{8}$~~~  \\
~~~[0.20, 0.25]~~~ & ~~~$5.63 \times 10^{8}$~~~  \\
~~~[0.25, 0.30]~~~ & ~~~$4.46 \times 10^{8}$~~~  \\
~~~[0.30, 0.35]~~~ & ~~~$3.66 \times 10^{8}$~~~  \\
~~~[0.35, 0.40]~~~ & ~~~$3.08 \times 10^{8}$~~~  \\
~~~[0.40, 0.45]~~~ & ~~~$2.66 \times 10^{8}$~~~  \\
\hline\hline
\end{tabular}
\end{table}

\subsection{Estimation of events}
\label{sec:events}

The signal events of the reaction between photons and the neutrons in the beam pipe can be estimated using \cite{Wang:2024ikx}
\beq\label{eq:Nsig}
  N^{\textrm{sig}} = N_n \times \sum_{E_{\gamma,\textrm{min.}}}^{E_{\gamma,\textrm{max.}}} \Big[ \sigma_{\textrm{bound}}(E_\gamma) \times \mathcal{L}_{\textrm{eff}}(E_\gamma) \Big] \,,
\eeq
where $E_{\gamma,\textrm{min.}} = 0.10\,\textrm{GeV}$ and $E_{\gamma,\textrm{max.}} = 0.45\,\textrm{GeV}$. Here $N_n = 5$ is the number of neutrons in beryllium. The cross section $\sigma_{\textrm{bound}}(E_\gamma)$ has been given in Table~\ref{tab:CSvalues}. The effective differential luminosity $\mathcal{L}_{\textrm{eff}}$ of the photon flux and target materials is defined as
\beq\label{eq:Leff}
  \mathcal{L}_{\textrm{eff}}(E_\gamma) = \mathcal{I}_{\textrm{Bha.}}(E_\gamma) \times \int_a^b N(x) \, dx \,,
\eeq
with
\beq\label{eq:Ibhabha}
  \mathcal{I}_{\textrm{Bha.}}(E_\gamma) = \int_{E_\gamma - \Delta E}^{E_\gamma + \Delta E} \rho_{\textrm{Bha.}}(E_\gamma') \, dE_\gamma' \,.
\eeq
Here $\rho_{\textrm{Bha.}}(E_\gamma')$ is the differential energy distribution of photon numbers. As indicated in Table~\ref{tab:rhodistribution}, the energy bin width is $\Delta E = 0.05\,\textrm{GeV}$.

The beam pipe consists of multiple layers of composite material, predominantly composed of beryllium. $N(x)$ is the number of nuclei per unit volume, and $a = 3.15\,\textrm{cm}$ and $b = 3.37\,\textrm{cm}$ are the distances from the inner surface and outer surface of the beam pipe to the $e^+ e^-$ collision axis. By integrating the material density along the beam pipe thickness, we obtain \cite{BESIII:2009fln}
\beq\label{eq:Ndensity}
  \int_a^b N(x) \, dx = 2.1 \times 10^{22} \,\textrm{cm}^{-2} \,.
\eeq
Combining all these results, the total number of $\gamma n \to \pi^- p$ events produced on the beryllium beam pipe at BESIII is estimated to be:
\beq\label{eq:result}
  N^{\textrm{sig}} \approx 81889 \,.
\eeq

This remarkably large number of events demonstrates the exceptional feasibility of studying the $\gamma n \to \pi^- p$ process at BESIII. The statistical uncertainty can be reduced to the percent level or below, allowing BESIII to precisely measure the cross section shown in Fig.\,\ref{fig:CSBeryllium}. By comparing the difference between $\sigma_{\textrm{bound}}(E_\gamma)$ and $\sigma_{\textrm{free}}(E_\gamma)$, one can quantitatively extract the nuclear modification effects. Furthermore, by searching for high-momentum initial-state neutrons among these events, one can precisely extract the properties of neutrons in SRC pairs within the beryllium nucleus---such as missing energy and effective mass. These measurements will address a class of experiments for which data are currently very scarce, and thus provide crucial information for a deeper understanding of nuclear structure.

It is worth noting that BESIII has now collected data samples with an integrated luminosity of approximately $37.65\,\textrm{fb}^{-1}$ above $3.5\,\textrm{GeV}$. This larger dataset yields roughly a factor of two increase in the number of signal events compared to our baseline estimate. Moreover, the Super $\tau$-Charm Facility (STCF) \cite{Achasov:2023gey} will reach a peak luminosity of $0.5 \times 10^{35}\,\textrm{cm}^{-2}\textrm{s}^{-1}$, which is 50 times that of the present BEPCII. This will further enhance the statistics by two orders of magnitude, making the STCF an exceptionally powerful machine for investigating the neutron momentum distribution through this process.

\section{Summary}
\label{sec:summary}
Atomic nuclei are finite quantum many-body systems whose structure emerges from the interplay of protons and neutrons bound by the strong interaction. A central goal of nuclear physics is to characterize this structure across multiple length and momentum scales, from the spatial extent of the nucleus to the short-distance dynamics of individual nucleons and nucleon pairs. In this context, nucleon momentum distributions provide a key link between nuclear structure and reaction dynamics, and are indispensable for the quantitative interpretation of experiments involving nuclear targets. In particular, the neutron momentum distribution plays a critical role: it is indispensable for elucidating the isospin structure of nucleon-nucleon correlations.

In this work, we have proposed to investigate the neutron momentum distribution in nuclei through the $\gamma n \to \pi^- p$ process at an electron-positron collider. We have demonstrated that photons produced by radiative Bhabha at BESIII can interact with the beryllium beam pipe, providing an ideal platform for this study. Within the impulse approximation, we have calculated the bound cross section for both deuteron and beryllium targets. For the deuteron, our predictions align with the existing PIONS@MAX-lab data near threshold. For the beryllium, the nuclear modification of the cross section is substantial, with the high-momentum neutrons in SRC pairs providing a pronounced contribution. We have estimated that approximately $8 \times 10^4$ signal events can be produced at BESIII, enabling percent-level precision measurements of the cross section shape as a function of photon energy. The proposal in this work can help to elucidate the structure of the nucleus and enhance our understanding of fundamental aspects of QCD. We encourage our experimental colleagues to pursue this investigation.

\section*{Acknowledgements}
The authors would like to thank Profs.\,Guang-Shun Huang, Shuang-Shi Fang, Wen-Biao Yan and Xiao-Rong Zhou for inspiring discussions and suggestions on the measurements at BESIII. This work is supported in part by National Key R\&D Program of China under Contracts No. 2023YFA1609400, and by the National Natural Science Foundation of China under Grant Nos.~12335003, 12535005, 12475098, 12105247, 12205255, and by the Fundamental Research Funds for the Central Universities under No. lzujbky-2023-stlt01, lzujbky-2024-oy02 and lzujbky-2025-eyt01, the Scientific Research Innovation Capability Support Project for Young Faculty under Grant No. ZYGXQNJSKYCXNLZCXM-P2. J.X is supported in part by the Key Laboratory for Particle Astrophysics and Cosmology, Ministry of Education (MoE).

\end{document}